\documentclass[epj]{svjour}

\usepackage{url}
\usepackage{natbib,graphicx}
\usepackage{float,pst-all}
\usepackage[english]{babel}
\usepackage{latexsym}
\usepackage{amssymb}
\usepackage{amsmath}

\newcommand{\be}{\begin{equation}}
\newcommand{\ee}{\end{equation}}

\begin{document}

\title{Electric and magnetic response to the continuum for $A=7$ isobars in a dicluster model}
\titlerunning{Response to the continuum for $A=7$ isobars}
\author{A.Mason, R.Chatterjee, L.Fortunato and A.Vitturi}
\authorrunning{A.Mason, {\it et al.}}
\institute{Dipartimento di Fisica "G.Galilei", Universit\`a di Padova and INFN,\\
v.Marzolo,8, I-35131 Padova, Italy}

\abstract{
Mirror isobars $^7$Li and $^7$Be are investigated in a dicluster model. The magnetic dipole moments and 
the magnetic dipole response to the continuum are calculated in this framework.
The magnetic contribution is found to be small with respect to electric dipole and quadrupole excitations 
even at astrophysical energies, at a variance with the case of deuteron.
Energy weighted molecular sum rules are evaluated and a formula for the molecular magnetic dipole sum rule 
is found which matches the numerical calculations.  
Cross-sections for photo-dissociation and radiative capture as well as  
the S-factor for reactions of astrophysical significance are calculated with good agreement with known
experimental data.} 
\maketitle

\section{Dicluster description of $^7$Be and $^7$Li}

It is well known that the properties of $A=7$ mass nuclei may be effectively described 
in terms of a dicluster model: two (inert) clusters in interaction with each other \cite{WalFli,WiTa, Buck,Kaj1,
Kaj2,Kaj3,Alt,Lan}. 
A long-standing description of these nuclei with such a dicluster picture has achieved excellent results
and a continued interest because it catches the essential physics within a very simple, yet powerful, model. 
Such a simple nuclear structure scheme may then used in reaction calculations \cite{Forvit}.
The need to go beyond this level of description may be justified only with the aim of taking into account more
refined considerations on the role of the Pauli principle or on the occurrence of complex phenomena like 
polarization of the clusters, and not because the model itself does not reproduce static and dynamic properties of 
$A=7$ nuclei to a good degree of accuracy. For example {\it ab initio} shell model calculations using a 
computationally-heavy Monte Carlo method have successfully reproduced many properties of these and other light 
nuclei starting from a nucleon-nucleon interaction \cite{Noll}. By computing ground state overlap functions, 
these calculations have shown that $^7$Be has a spectroscopic factor of almost 1 in the 
$\alpha+\;^3$He configuration and, similarly, $^7$Li has a large spectroscopic factor in the $\alpha+t$ channel.
A number of other approaches, that we will not discuss here, such as Resonating Group Methods or Generator 
Coordinates Methods, are available to various degrees of complication and success. 
In addition we must mention a very recent study by Canton and Levchuk on the low-energy capture reactions \cite{Cant} 
using the Multichannel Algebraic Scattering approach that has been applied to the $^3$He($\alpha,\gamma$)$^7$Be 
reaction at astrophysical energies achieving a good agreement with S-factors data.

We have shown in recent papers \cite{Forvit} that the dicluster model
may be profitably used to calculate a number of static and dynamic properties of $^7$Li, especially in connection with
excitations to the continuum in break-up processes. We apply here the same model
to $^7$Be and in addition we investigate the magnetic excitations to the continuum that were neglected in 
previous works. It is in fact important to give proper estimates of 
their contributions to reaction cross-sections, especially at
energies of astrophysical interest. 
These two isospin mirrors not only display very similar ground state properties, but also manifest a similar
behaviour as long as their response to the continuum is considered. Nowadays it is becoming more and more evident 
that to properly describe structure and reactions of loosely bound systems, e.g. nuclei along the drip lines, 
it is essential to take into account the coupling to the continuum and the break up channel. This is relatively simple 
in the case of one-particle halo nuclei. In the case of more complex weakly-bound nuclei such as  $^7$Li or $^7$Be,
the cluster picture offers the possibility of describing the ground and a few excited states in terms formally analogous 
to that of a single-particle picture, with the {\it caveat} that
the single particle is not just a neutron or a proton, but a composite one, 
like for instance a triton.

Of course, as in any theory driven by phenomenological considerations, a certain degree of approximation comes into 
play, and one needs to consider the pros and cons of the approach. We mention therefore that 
considering the clusters as spherical and elementary, but not point-like, has some implications:
i) one neglects possible polarization effect, which may come into play because of the proximity of the two 
clusters and ii) one neglects nucleon-exchange effects between the two clusters.
We shall prove that these limitations are not severe (see also \cite{Forvit}) and the model works rather well
for the $A=7$ nuclei.

Our intercluster potential is assumed to be the sum of a nuclear Woods-Saxon potential, 
a Coulomb potential and a spin-orbit interaction:
\begin{equation} \label{pot}
V_{A_1-A_2} = V_{coul}(r)+ V_{WS}(r) + 
V_{{\bf l}\cdot {\bf s}}(r)\;.
\end{equation}
with $V_{{\bf l}\cdot {\bf s}}(r)= V_{{\bf l}\cdot {\bf s}}{\bf l}\cdot {\bf s} \frac{r_0^2}{r} \frac{dV_{WS}(r)}{dr}$. 
Here $r_0$ is the same radius parameter that enters into the parametrization of the Woods-Saxon potential, $\vec \ell$ 
and $\vec s$ are the orbital angular momentum and spin of the $^3$H($^3$He) cluster in $^7$Li($^7$Be) respectively. 
The Coulomb potential has the standard $1/r$ behaviour at large distances (point charge), 
but at small distances we use the Coulomb potential generated by a uniform spherical charge distribution.

Note that it is not possible to identify one of the two clusters as a core and the other as 
a valence particle, since the two fragments have comparable masses and charges. In the 
spin-orbit interaction, $r_0$ is the radius of the Woods-Saxon potential, 
$\bf l$ refers to the relative angular momentum, and $\bf s$ refers 
to the spin of the triton (helium-3) which can be seen as a ``heavy" single-particle for  
$^7$Li ($^7$Be).

The $A=7$ isobars are particularly challenging.
They are weakly bound, but nevertheless they have one excited bound state. 
The low-lying response presents (as we will show) non-resonant as well as resonant peaks with comparable 
intensities. Aside from the importance in pure nuclear physics, the break-up, photo-dissociation and 
capture processes for these nuclei play an important role for astrophysics and for applicative purposes 
(e.g. lithium is used as coolant in nuclear reactors). We apply, in particular, our dicluster model
 to the calculation of astrophysical S-factors for the reaction
$^3$He($\alpha,\gamma$)$^7$Be (and also $^3$H($\alpha,\gamma$)$^7$Li). Recently the LUNA 
collaboration \cite{Luna} has undergone a successful experimental campaign aimed at measuring the cross-section
for this reaction which is thought to i) strongly influence the hydrogen-burning process in the Sun and 
consequently the solar neutrino production and ii) determine the primordial abundance of light elements 
which influences the determination of the baryon-to-photon ratio of the Universe from comparison with nucleosynthesis. In both these fields the determination of this 
quantity is crucial and this fact justifies even more the need for simple, but yet powerful, 
theories that 
could explain and reproduce the data. For a review of experiments in this direction see Ref. \cite{Luna}.

The paper is organized as follows: the present section contains a summary of 
the method and deals with the parameters of the cluster-cluster potential 
and with the electromagnetic properties of the $^7$Li and $^7$Be systems. Static properties and transition rates among
bound states are compared with data whenever possible. Section II contains a digression 
about molecular sum rules, their evaluation in the present case and the derivation of a new energy weighted sum rule 
for magnetic dipole transitions, inspired to the Kurath single-particle sum rule\cite{Kur,Trai}. 
Section III deals with the application to 
astrophysics and presents our results for the photo-dissociation and radiative capture reactions. In Section IV 
we present the S-factors for the two reactions $^3$He($\alpha,\gamma$)$^7$Be and $^3$H($\alpha,\gamma$)$^7$Li in 
comparison with known experimental measurements. The conclusions are followed by an appendix where a few useful 
formulas for dicluster nuclei are derived.

\subsection{$^7$Be}

We consider  $^7$Be as formed by $\alpha$ and $^3$He. 
The parameters of the cluster-cluster Woods-Saxon potential are
 $V_{WS}=-73.851$ MeV for the depth, $1.60$ fm and
$0.48$ fm for its radius and diffusivity. The magnitude of the spin-orbit 
potential, $V_{{\bf l}\cdot {\bf s}}=1.275$ MeV, is chosen in order to reproduce the 
splitting of the two bound eigenstates, respectively at $-1.58$ and $-0.98$ MeV. 
The shape of the potential has been tuned in order to obtain an overall
agreement with various observables, like the charge and mass radii.
The $\alpha$-particle 
has an intrinsic spin equal to zero, $^3$He has spin $1/2$ and the 
intercluster orbital angular momentum for the ground-state is $\ell=1$: this can either be postulated on
the basis of the Wildermuth connection \cite{WiTa} or it can be inferred 
phenomenologically, for example, from the magnetic moment for a dicluster nucleus.
Using the intrinsic magnetic moment, masses and charges 
of the two clusters in formula (\ref{muAB}) for the total magnetic moment, 
one can see from Table (\ref{mucompare}) that the value that gives the best agreement is clearly $\ell=1$ (which also gives the correct parity).
Due to spin-orbit interaction the ground state has $j=3/2^{-}$, while the first excited state has 
$j=1/2^-$. 

\begin{table}[htbp]
\begin{center}
\begin{tabular}{|c|c|c|c||c|}
\hline
&s&p&d&exp.\\ \hline
 $^7$Li&2.97&3.37&3.78&3.256427(2)\\
 $^7$Be&-2.12&-1.53&-0.93&-1.398(15)\\ \hline
\end{tabular}
\end{center}
\caption{Total magnetic moments (in $\mu_N$) of  $^7$Be and $^7$Li for different choices of relative angular
 momenta compared with experimental value (last column). 
Even forgetting parity considerations that would of course leave only the $\ell=1$ state as a possible candidate, 
it is clear that the best agreement is obtained in both cases by considering a relative p-wave. } 
\label{mucompare}
\end{table}

We calculate the relative motion wavefunction of the ground state, first excited state 
and continuum states by solving numerically the radial Schr\"odinger 
equation with the intercluster potential given in (\ref{pot}). In order to 
yield the resonant $7/2^-$ and $5/2^-$ states, we have to use a different set of 
parameters for the Woods-Saxon ($V_{WS}=-64.68$ MeV) and spin-orbit 
(V$_{ls}=1.94$ MeV) potentials, but keeping the same geometry. 
With this choice we find that the width of the 
$f_{7/2}$ and $f_{5/2}$ resonances are in reasonable agreement with the experimental data 
(see table (\ref{static})). A number of ground state properties are 
found to be in good agreement with measured observables.
We also calculate the $B(E2, 3/2^- \rightarrow 1/2^-)$ and 
$B(M1,3/2^-\rightarrow 1/2^-)$ transition strengths between the ground state 
and the first excited bound state.
All the calculated observables, obtained using the expressions given in appendix, 
are collected in table (\ref{static}) with a 
comparison with known experimental data. 

\begin{table}[t]
\begin{center}
\begin{tabular}{|l|c|c|}
\hline
Quantity & this work & experiments \\
\hline  \hline   
$\langle r^2\rangle_{A+B}$ (fm) & 2.48  & 2.48 $\pm$ 0.03 \\
\hline
$ \langle r^2 \rangle_{A+B}^{ch}$ (fm)  & 2.52  & 2.52 $\pm$ 0.03 \\
\hline

$Q^{matter}$ (fm$^2$) & -11.83 & - \\
\hline
$Q^{charge}$ (fm$^2$) & - 4.79  & - \\
\hline
$\mu$ ($\mu_N$)   &    -1.53    & -1.398 $\pm$ 0.015 \\
\hline
$B(E2, 3/2^-\rightarrow 1/2^-)$ ($e^2fm^4$)  & 18.3 & -  \\
\hline
 $B(M1,3/2^-\rightarrow 1/2^-)$ ($\mu^2$)& 1.86 &  1.87 $\pm$ 0.25 \\
\hline
$\Gamma(7/2^-)$ (keV)  & $\sim$ 90   &   175 $\pm$ 7 \\
\hline
 $\Gamma(5/2^-)$ (MeV) & $\sim$ 0.8  &   1.2  \\
\hline 
\end{tabular}
\end{center}
\caption{Static properties of $^7$Be in a dicluster model and electromagnetic transition strengths
between bound states. Definitions of some of these observables can be found 
in the appendix. Experimental values are taken from \cite{Ta,1998,Liu,Hu}}
\label{static}
\end{table}

\begin{figure}[t!]
\begin{center}
\includegraphics[clip=,width=.48\textwidth]{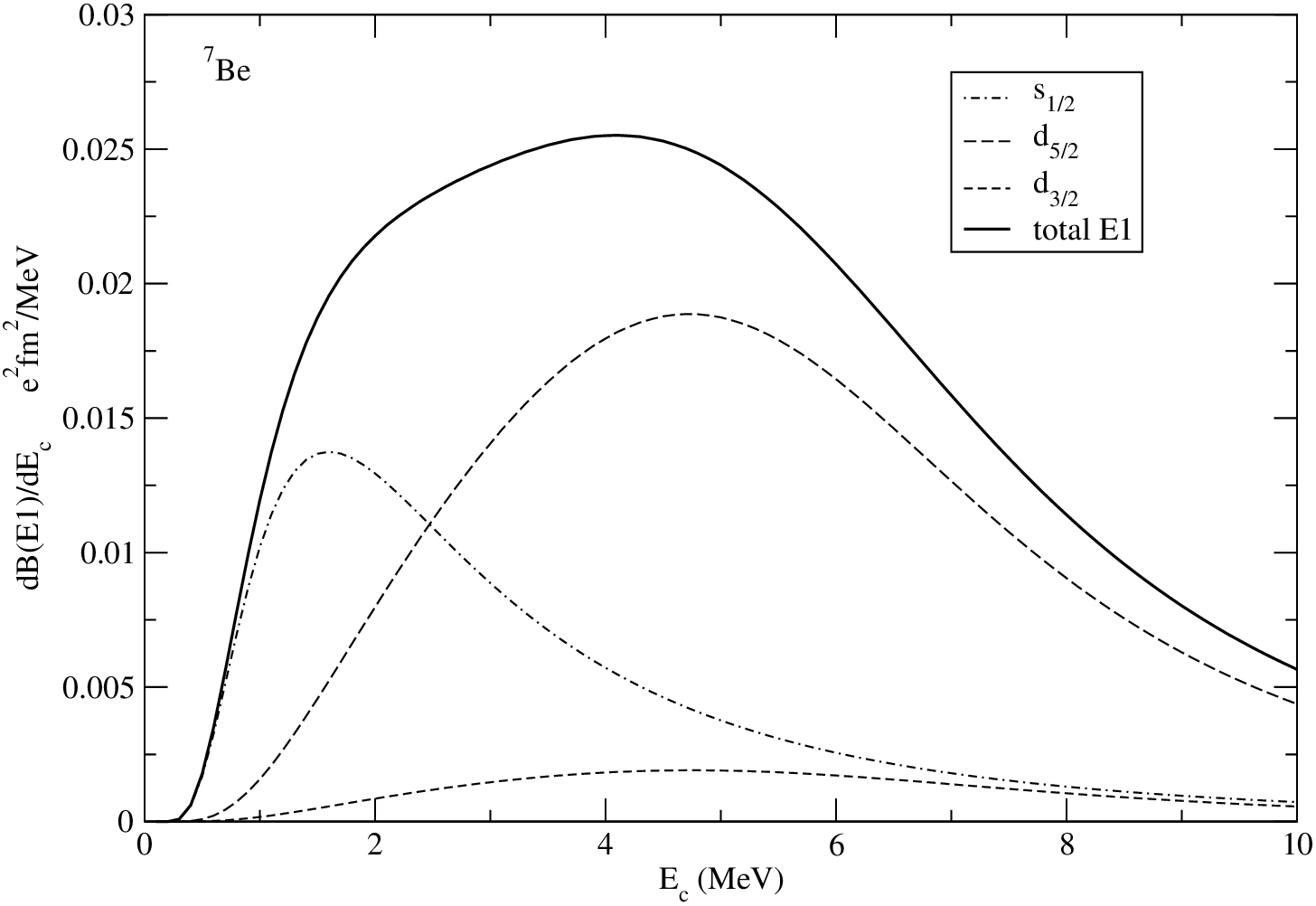}
\end{center}
\caption{Differential dB(E1)/dE$_c$ in e$^2$fm$^2$/MeV for transitions 
from the ground-state to the continuum. Energies are in MeV, referred to 
the threshold for breaking  into the $\alpha-^3$He channel.
Separated contributions are indicated in the legend.
Solid line shows the sum of all the electric dipole contributions.}
\label{BeE1}
\end{figure}

As it has been done in Ref. \cite{Forvit} for $^7$Li, besides the quadrupole transition to the first 
excited bound state given in Table (\ref{static}),  we also investigate the electromagnetic response 
of $^7$Be to continuum states.
Figure (\ref{BeE1}) shows differential reduced transition probability (see Ref.\cite{tb}), $dB(E1)/dE_c$, for 
continuum states allowed by an E1 transition. 
In this case there are no resonances in the low-lying continuum and the 
visible peaks have a non-resonant nature \cite{Cata}.
Dotted line refers to continuum $s_{1/2}$ state, dashed line and long-dashed refers to $d_{3/2}$ 
and $d_{5/2}$ states respectively.  
The solid line gives the sum of all contributions.
The total integrated non-energy weighted B(E1) is 0.226 e$^2$fm$^2$ (See next section for comments on sum rules).  

\begin{figure}[t!]
\begin{center}
\includegraphics[clip=,width=.48\textwidth]{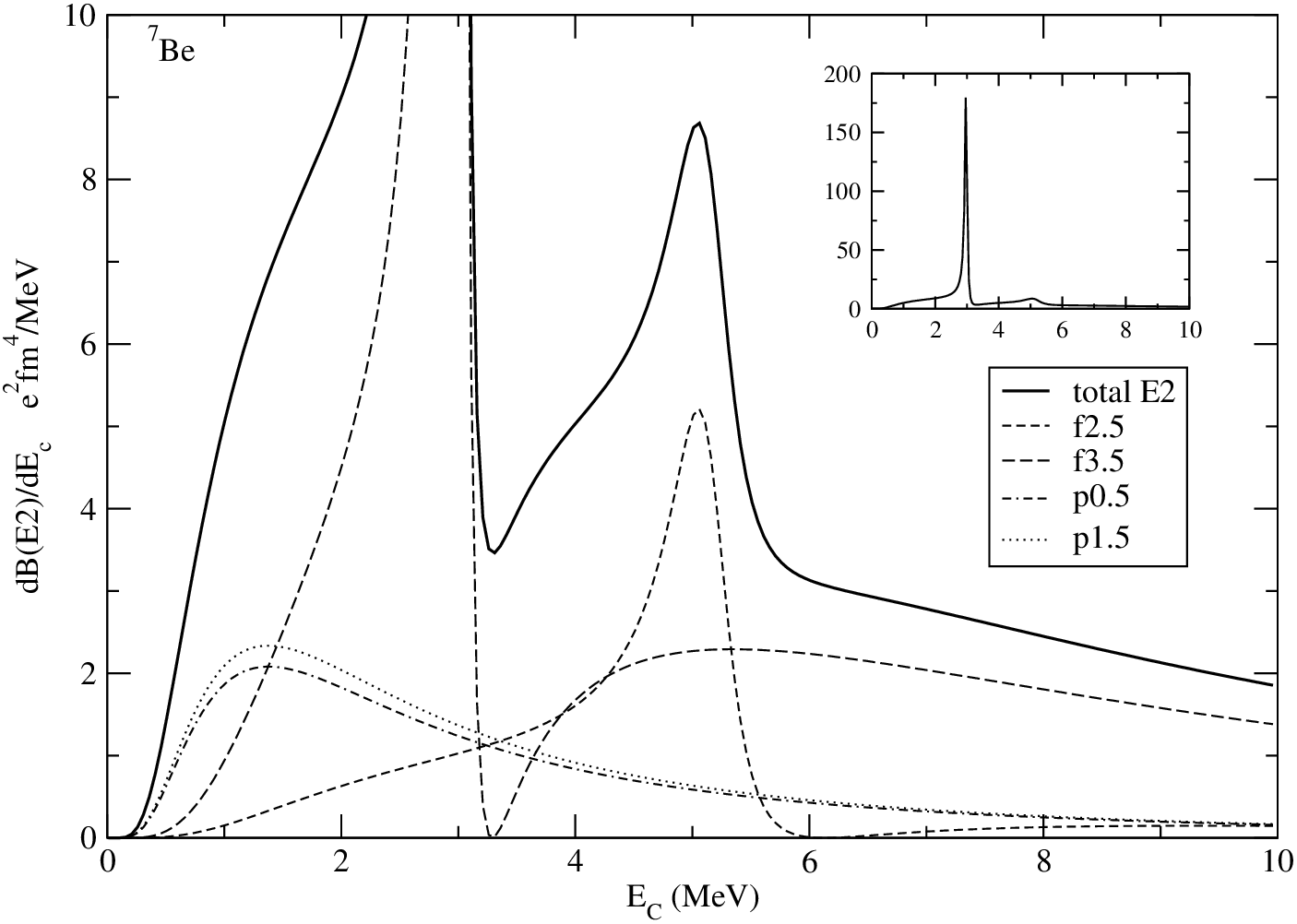}
\end{center}
\caption{ Differential dB(E2)/E$_c$ in e$^2$fm$^4$/MeV for transitions 
from the ground-state to the continuum. Energies are in MeV, referred to 
the threshold for breaking  into the $\alpha-^3$He channel. Separated 
contributions are indicated in the legend.
Solid line shows the sum of all the electric quadrupole contributions. }
\label{BeE2}
\end{figure}

Figure (\ref{BeE2})  shows differential $dB(E2)/dE_c$ for 
continuum states allowed  by an E2 transition. Long-dashed and 
dot-dashed lines refer to $f_{5/2}$ and $f_{7/2}$ states respectively, 
while dotted and dashed lines refer to $p_{1/2}$ and $p_{3/2}$ states respectively.
The $f$ states take contributions both from the resonances and from the non-resonant part of the continuum,
while the peaks of the $p$ states arise purely from non resonant transitions. 
We show also in the inset the full $f_{7/2}$ resonance
on a different scale because it reaches a maximum of about 150 e$^2$fm$^4$/MeV.  
Finally solid line gives the sum of all contributions.
The total integrated non-energy weighted B(E2) is 113.5 e$^2$fm$^4$ to be compared with
the $B(E2,  3/2^-\rightarrow 1/2^-)$ value of 18.3 e$^2$fm$^4$. 
Non resonant continuum states $p_{1/2}$ and $p_{3/2}$ are 
dominant at very low continuum energy, while they become
negligible at higher energies around and beyond the $f_{5/2}$ peak.

In addition to electric excitations we also consider magnetic ones.
Magnetic dipole interactions give normally contributions to the transition probabilities 
that are almost comparable with electric quadrupole and smaller than electric dipole. 
It appears however, already in old calculations \cite{BlWei} for the photo-disintegration 
of the deuteron (the smallest dicluster nucleus), that they are relevant at 
astrophysical energies (say below 200 keV) where they even turn out to dominate over the electric dipole.
For this reason we decided to investigate this theme to better understand the role
they might play in low-energy nuclear reactions involving dicluster systems (cfr Sect. 4).

The expression for the magnetic operator can be obtained from the multipoles expansions of 
the electromagnetic field with sources as (cfr. Ref. \cite{DST}, ch.17):
\begin{equation}
{\it M}(M1)=\sum_i[\nabla(r_i^\lambda Y_{\lambda}(\theta,\phi))]\cdot
\bigl[\frac{{e^\lambda_{eff}}_i\hbar}{m_i ec(\lambda+1)} \vec \ell_i + \frac{\mu_i \vec \sigma_i  }{e} 
\bigr]
\end{equation}
where the sum is over all particles (with cordinates $r,\theta,\phi$), 
each with an effective charge ${e^\lambda_{eff}}_i$, magnetic moment $\mu_i$ and mass $m_i$.
As usual $\lambda$ indicate the multipolarity, $e$ is the electron charge.

In the present case the wave functions are not expressed in a single-particle proton or neutron basis, 
but rather in a two-cluster basis with the $\alpha$ (which is spinless) and either the triton or the helium-3. 
Therefore we cannot adopt the same simplifications as in
Ref. \cite{DST}, but we have to rewrite the magnetic operator in a cluster form. 
For the dipole case this is not difficult and the magnetic operator can be recast in the form:
$${\it M}(M1)=\sqrt{3} \vec Y_0 \times(\vec \mu_1 +\vec \mu_2+ \mu_N G\vec L) = $$
\begin{equation}
= \sqrt{3} \vec Y_0 \times(\mu_N G\vec J+
(2\mu_{cl}-G)\vec S)  
\end{equation}
where the summation over the $A$ particles has been simplified by partitioning the protons and neutrons into 
two spherical clusters (see appendix). Here $\bf Y$ indicates vector spherical harmonics, 
$\mu_1$ and $\mu_2$ are the intrinsic static magnetic operator of the clusters (in nuclear magnetons, $\mu_N$)
and $\vec L=\vec J-\vec S$ is the cluster-cluster orbital angular momentum operator. The total angular momentum and spin of the cluster are indicated with $\bf J$ and $\bf S$ respectively. The coefficient 
$G=\frac{Z_1A_2^2+Z_2A_1^2}{AA_1A_2}$ acts as an effective cluster-cluster orbital gyromagnetic factor.
Since the $\alpha$ has null spin and null magnetic moment, 
only the intrinsic magnetic moment of the other cluster appears ($\mu_{cl}=\mu_2$).

\begin{figure}[t!]
\begin{center}
\includegraphics[clip=,width=.48\textwidth]{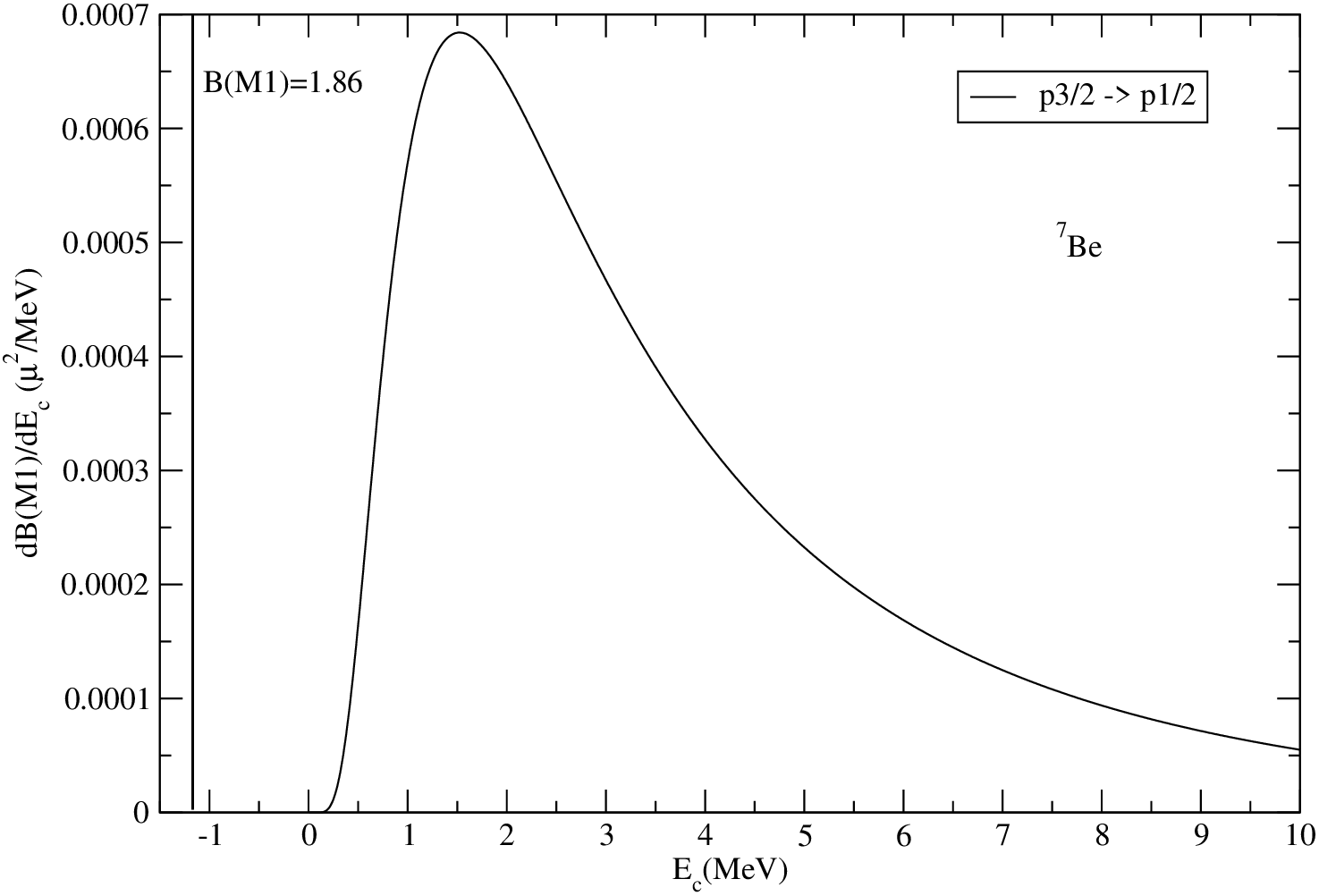}
\end{center}
\caption{Differential  dB(M1)/dE$_c$in $\mu^2$/MeV for transitions 
from ground-state to the continuum. Energies are in MeV, referred to 
the threshold for breaking  into the $\alpha-^3$He channel.}
\label{BeM1}
\end{figure}

By calculating the reduced matrix elements along the lines of Ref. \cite{DST}, one gets:
$$B(M1)= 9 (2j_f+1) \mid \langle l_f j_f \mid l_i,j_i \rangle \mid^2 \langle \ell_f ||\vec Y_0 ||\ell_i\rangle^2 \cdot$$
$$\Biggl[ \mu_N G(-1)^{\ell_f+s_f+j_f-1}\langle \ell_i j_i || \vec J ||\ell_i,j_i\rangle 
\left\{ {j_f \atop 1} {1 \atop j_i} {j_i \atop 0} \right\} \left\{ {l_f \atop j_i} {j_f \atop l_i} {s_f \atop 0} \right\}  $$ 
\begin{equation}
+ (2\mu_{cl}-\mu_N G)\langle s_f||\vec s|| s_i\rangle \begin{Bmatrix}
l_f & s_f & j_f\\
l_i & s_i & j_i\\
0 & 1 & 1 
\end{Bmatrix} \Biggr]
\end{equation}
where $l_i,j_i,l_f$ and $j_f$ are initial and final orbital and total angular momenta.
The selection rule given by the overlap and the reduced matrix element of the vector spherical harmonic
imply that one can only populate states with $\ell_i=\ell_f$ and $j_i \ne j_f$.
In our specific case, where the initial state has $\ell_j=p_{3/2}$, we can only reach states with
$p_{1/2}$ character. The corresponding B(M1) is :
\begin{equation}
B(M1)=   {1\over 4\pi} (2\mu_{cl}-\mu_N G)^2 \mid \langle l_f j_f \mid l_i,j_i \rangle \mid^2
\end{equation}
that coincides to the formula given in Refs. \cite{WalFli,Buck}. 
The calculation of the $B(M1; 3/2 \rightarrow 1/2)$ to the bound state (that is within the two 
isospin partner levels) gives an excellent agreement with experimental measurements as seen from table \ref{static}.
Figure (\ref{BeM1}) shows continuum states excited in M1 transitions 
for $^7$Be. Only transitions to the $p_{1/2}$ states are allowed 
because of orthogonality between the ground-state $p_{3/2}$ and 
continuum $p_{3/2}$ (see the next section for a more detailed discussion).
The total non-energy weighted B(M1) to the continuum states, integrated 
up to 50 MeV, is about $0.00408~\mu^2$, much smaller than the magnetic dipole 
strength to the first bound state.

\subsection{$^7$Li}

In a the previous paper \cite{Forvit} we have investigated  
dipole and quadrupole electric transitions of the nucleus $^7$Li.
We refer here to the same parameter sets for the potential and we improve our 
calculations by adding results for magnetic dipole transitions to the continuum
 for the sake of completeness.
We note that $^7$Li and $^7$Be have the same qualitative 
behaviour because of their similar internal structure.  
An estimate of static magnetic dipole moment of $^7$Li ground-state obtained  
by using formula (\ref{muAB}) gives : $\mu(^7Li) = 3.37 \mu_N$ that is in good agreement
with the experimental value $3.256 \mu_N$ given in \cite{Aj}. Again this fact confirms 
that in a simplified dicluster picture the $\ell=1$ relative angular momentum gives 
a good magnetic dipole moment as in $^7$Be.

\begin{figure}[t!]
\begin{center}
\includegraphics[clip=,width=.48\textwidth]{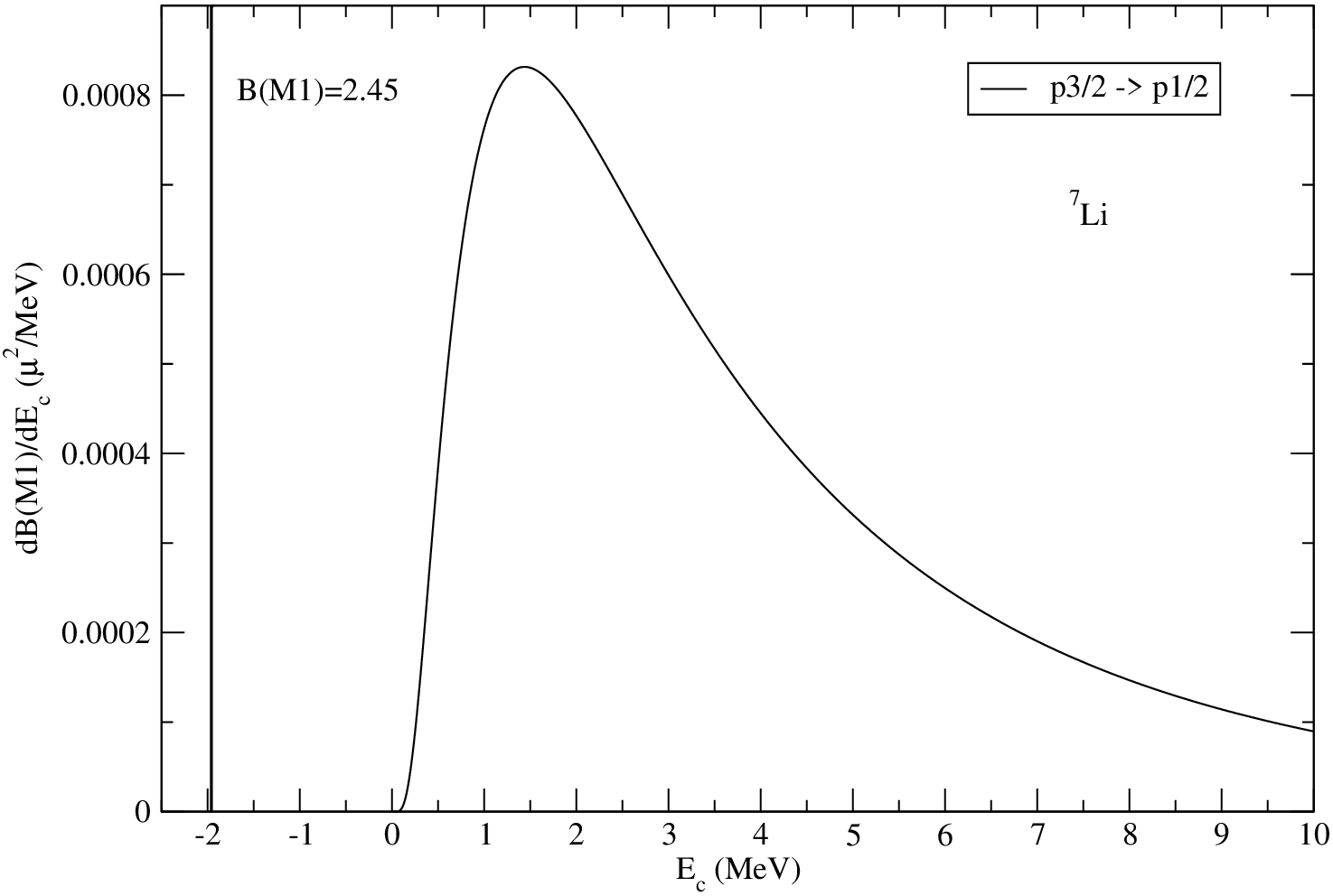}
\end{center}
\caption{ Differential dB(M1)/dE$_c$ in $\mu^2/MeV$ for transitions 
from ground-state to continuum. Energies are in MeV, referred to 
the threshold for breaking  into the $\alpha-t$ channel.}
\label{LiM1}
\end{figure}

$^7$Li and $^7$Be have also the same qualitative behaviour with respect to transitions to the 
continuum: compare plots in Ref. \cite{Forvit} with figures (\ref{BeE1}) and (\ref{BeE2}) in the present 
paper and compare figure (\ref{BeM1}) with  figure (\ref{LiM1}).
In particular the last one shows the magnetic dipole transition strength for transitions between the 
ground-state and the continuum. Only transition to $p_{1/2}$ continuum states are allowed 
because of orthogonality between the ground-state $p_{3/2}$ and the $p_{3/2}$ continuum. 
This point was discussed, for another system, also in Ref. \cite{Kro}.

The total integrated $B(M1)$ strength amounts to $0.0029 \mu^2$ again much smaller 
than the B(M1) to
the first excited bound state. 
The magnetic dipole contribution is, in both nuclides, not as important as for the deuteron.

\section{Molecular sum rules}
Useful tools in nuclear physics are sum rules \cite{Bohr,alga,Lan,Ber}.
In particular energy weighted sum rules (EWSM)
 for an electromagnetic interaction can be expressed in terms of proper expectation values in the ground state
$$ S(\pi\lambda) \equiv \sum_f (E_f-E_i)B(\pi\lambda;i \rightarrow f)=$$
\begin{equation}  \label{def}
={1\over 2} 
\langle 0\mid [\hat O(\pi\lambda),[H,\hat O (\pi\lambda)]]\mid 0\rangle
\end{equation}
where $i$ stands for the initial state and the sum is extended over all reachable 
final states $f$. Here $H$ is the hamiltonian operator of the system. 
The energies of the initial and final states are $E_i$ and $E_f$ 
and the operator $\hat O (\pi\lambda)$ in the double commutator is the one that induces the 
transition. For electric dipole
and quadrupole interactions, equation (\ref{def}) leads respectively to \cite{Bohr}:
\begin{eqnarray} 
S(E1) &=& \frac{9}{4\pi} \frac{\hbar^2}{2m} \frac{NZ}{A}e^2 
\nonumber \\
S(E2) &=& \frac{50}{4\pi}\frac{\hbar^2}{2m}Ze^2\langle r^2 \rangle^{ch} \,.
\label{EWSR}
\end{eqnarray}
where $m$ is the nucleon mass and $\langle r^2 \rangle^{ch}$ is the mean square charge radius.
Our calculations show that the EWSR (\ref{EWSR}) are clearly not exhausted by considering only 
the transitions which imply changes in the relative motion (see table (\ref{tab2})). 
We assume that the suitable sum rules in this case 
are the energy weighted molecular sum rules (EWMSR) that can be obtained by removing 
the contributions of the individual clusters \cite{alga}. If a nucleus  with mass and charge 
(A, Z) is split into two clusters (A$_1$, Z$_1$) and (A$_2$, Z$_2$) then
the EWMSR is in general defined as
$$S_{mol}(E\lambda, A_1+A_2) =  $$
\begin{equation}    \label{mol}
=S(E\lambda, A_1+A_2) - S(E\lambda, A_1) - S
(E\lambda, A_2)  \,.
\end{equation}
For electric dipole \cite{alga} and quadrupole interactions one obtains: 
\begin{eqnarray*}     \label{mol2}
S_{mol}(E1) &=& \frac{9}{4\pi}\frac{(Z_1A_2 - Z_2A_1)^2}{AA_1A_2}
\frac{\hbar^2e^2}{2m} \nonumber \\
S_{mol}(E2) &=& \frac{25}{4\pi}\left(
Z_1\left(\frac{A_2}{A}\right)^2 + Z_2\left(\frac{A_1}{A}\right)^2
\right) \langle R\rangle^2 \frac{\hbar^2e^2}{m}
\end{eqnarray*}


The EWMSR for the quadrupole case is obtained by using formula (\ref{r2ch}) into eq. (\ref{mol}).
Note that in Ref. \cite{alga} the expectation value of  $\langle R\rangle^2$ was approximated, in a phenomenological
fashion, by the intercluster equilibrium distance, $S_0$.

Energy weighted sum rules for the magnetic dipole \cite{Kur} and higher multipolarities \cite{Trai} have
been evaluated in a pure single-particle scheme with a shell model hamiltonian containing a spin-orbit part.
The magnitude of the spin-orbit splitting is directly proportional to the total energy-weighted strength.
In our cluster picture, one can follow a similar idea to evaluate the Energy-Weighted Molecular Sum Rule
for the magnetic dipole interaction. The model hamiltonian is $\hat H=H_0+V_{so}(R)\vec L\cdot \vec S$, where $H_0$
contains the kinetic term and the central part of the potential, while the operator that
promotes magnetic dipole excitations amounts to the third component of $\hat O=\sum_{i=1,2}
(G\vec L_i+2\mu_{i}\vec S_i)$, where $G$ (defined above and also in \ref{muAB}) 
contains the combined effects of effective charges 
and reduced mass.
The expectation value in the ground state of double commutator gives
$$S_{mol}(M1,A_1+A_2)=\frac{1}{2}\langle 0\mid [\hat O_z,[\hat H,\hat O_z]]\mid 0\rangle = $$
\begin{equation}
=-\frac{1}{2}(G-2\mu_2)^2 \langle 0\mid V_{so}(R)\vec L \cdot \vec S\mid 0\rangle
\end{equation}

We have calculated all the contributions of E1 and E2 transitions, including
also the contribution of ``virtual'' states, or states that are not allowed by the Wildermuth's rule,
($0s_{1/2},1s_{1/2}, 0d_{5/2},0d_{3/2}$ for electric dipole and $0p_{3/2},
0p_{1/2}$ for quadrupole). For magnetic dipole interaction we have to consider 
 the virtual state $0p_{1/2}$, the first excited state $1p_{1/2}$ and the 
continuum $p_{1/2}$, but we do not need to consider the virtual state $0p_{3/2}$. 
The reason is the following: in the reduced matrix element for an M$\lambda$ 
interaction the radial contribution is $\langle f |r^{\lambda-1}|i\rangle$.
For magnetic dipole interaction this contribution is simply the overlap between 
the initial state, $i$, and the final state, $f$. If the states carry 
the same angular momentum and spin quantum numbers then their overlap is zero by orthogonality, 
because they are different eigenstates of the same hamiltonian (spin-orbit potential included).
For the same reason the virtual state  $0p_{1/2}$ does indeed contribute: since the spin-orbit part of the  
potential is different, the two hamiltonians are different and one cannot use the property of 
orthogonality, but rather has to calculate the actual overlap.

In table (\ref{tab2}) we compare EWSR and EWMSR with the values obtained in our model.
We find that the low-lying dipole strength exhausts almost entirely the EWMSM, but represents 
only a small fraction of the EWSR. In the same way the quadrupole strength coming from changes in
the relative motion of dicluster configurations exhausts a large fraction of the EWMSR, but only 
a small part of the standard EWSR.

\begin{table*}[htbp]
\begin{center}
\begin{tabular}{|c||c|c|c||c|c|c|}
 \hline
 & \multicolumn{3}{|c||}{$^7$Li} & \multicolumn{3}{c|}{$^7$Be} \\
 \hline
\hline
EM$\lambda$ & EWMSR  &  EWSR  & ours   & EWMSR  &  EWSR  & ours    \\
\hline
E1  & 1.02  &  36.7  &  1.01  &  1.02  &  36.7  & 1.00 \\
\hline
E2 & 1120 & 2105 &  481.6  & 1424 & 3018 & 639 \\
\hline \hline
M1 &  1.36  &  -   &    1.20 & 0.87 &  - &    0.82  \\
\hline
\end{tabular}
\end{center}
\caption{Comparison between EWSR, EWMSR and our calculations for $^7$Li and
$^7$Be. Values are in MeV$^2$fm$^3$ for E1, MeV$^2$fm$^5$ for E2
and $\mu_N^2$MeV for M1.}
\label{tab2}
\end{table*}

The energy weighted molecular sum rule for magnetic dipole transitions is fulfilled 
quite well by our calculations: 88\% in the case of lithium and 94\% in the case of berillium.
Practically all the contribution comes from the excitation to the first excited bound state.

\section{Photo-dissociation and radiative capture}

The knowledge of the electromagnetic response to continuum is a basic ingredient to describe 
break-up processes. In fact, in kinematic conditions where the process is dominated by the Coulomb interaction
(for example at very forward angles), the break-up probabilities become directly proportional to the $B(E\lambda)$
values. Independently from kinematic conditions this is also the case for two other processes
 of fundamental importance 
for astrophysics that can be therefore studied within the present model: the photo-dissociation and its inverse process, 
the radiative capture.

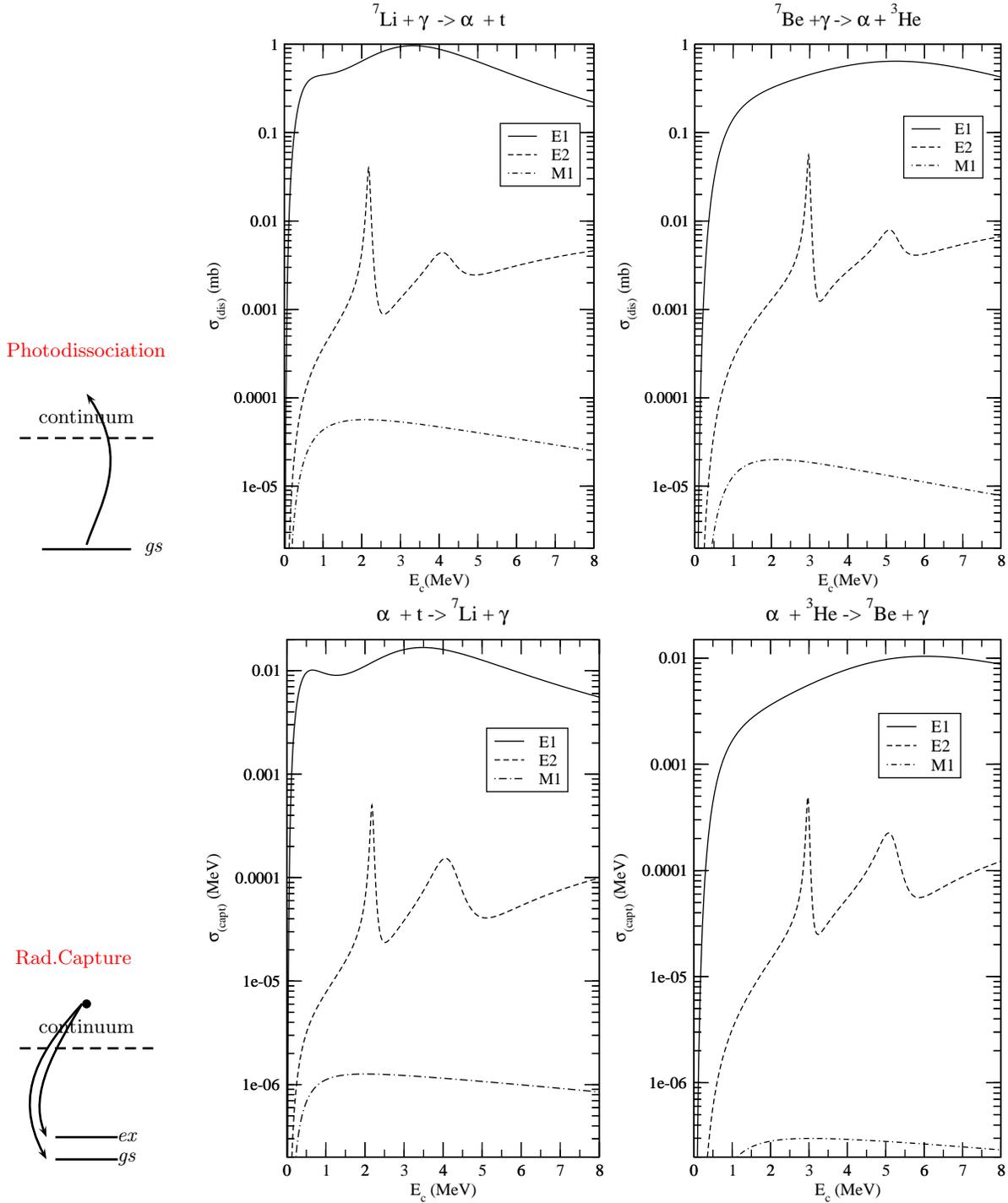
\begin{figure*}

\begin{picture}(100,230)(0,0)
\psset{linewidth=1pt,unit=0.07cm}
\psline(15,10)(35,10)
\psbezier{->}(25,11)(28,20)(35,30)(25,45)
\psline[linestyle=dashed](10,35)(40,35)
\rput(40,10){\em gs}
\rput(25,40){continuum}
\rput(25,55){\red Photodissociation}
\end{picture}
\includegraphics[clip=,width=0.70\textwidth]{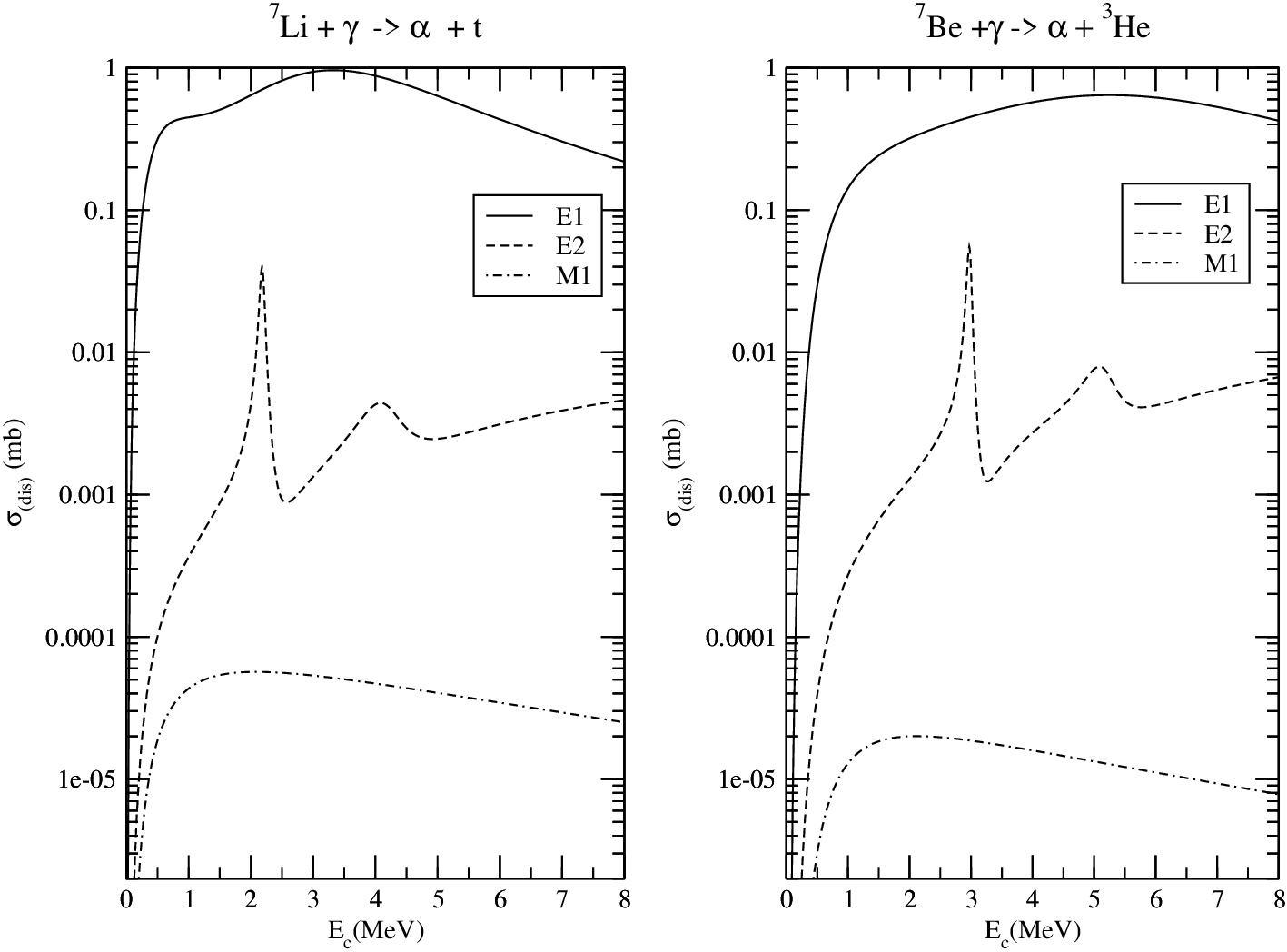}

\begin{picture}(100,230)(0,0)
\psset{linewidth=1pt,unit=0.07cm}
\psline[linestyle=dashed](10,35)(40,35)
\psline(18,15)(32,15)
\psline(18,10)(32,10)
\psdots*[dotscale=1,dotstyle=*](25,45)
\psbezier{->}(24,45)(18,35)(12,25)(16,15)
\psbezier{->}(24,45)(15,35)(8,25)(16,10)
\rput(34,10){\em gs}
\rput(34,15){\em ex}
\rput(22,55){\red Rad.Capture}
\rput(25,40){continuum}
\end{picture}
\includegraphics[clip=,width=0.70\textwidth]{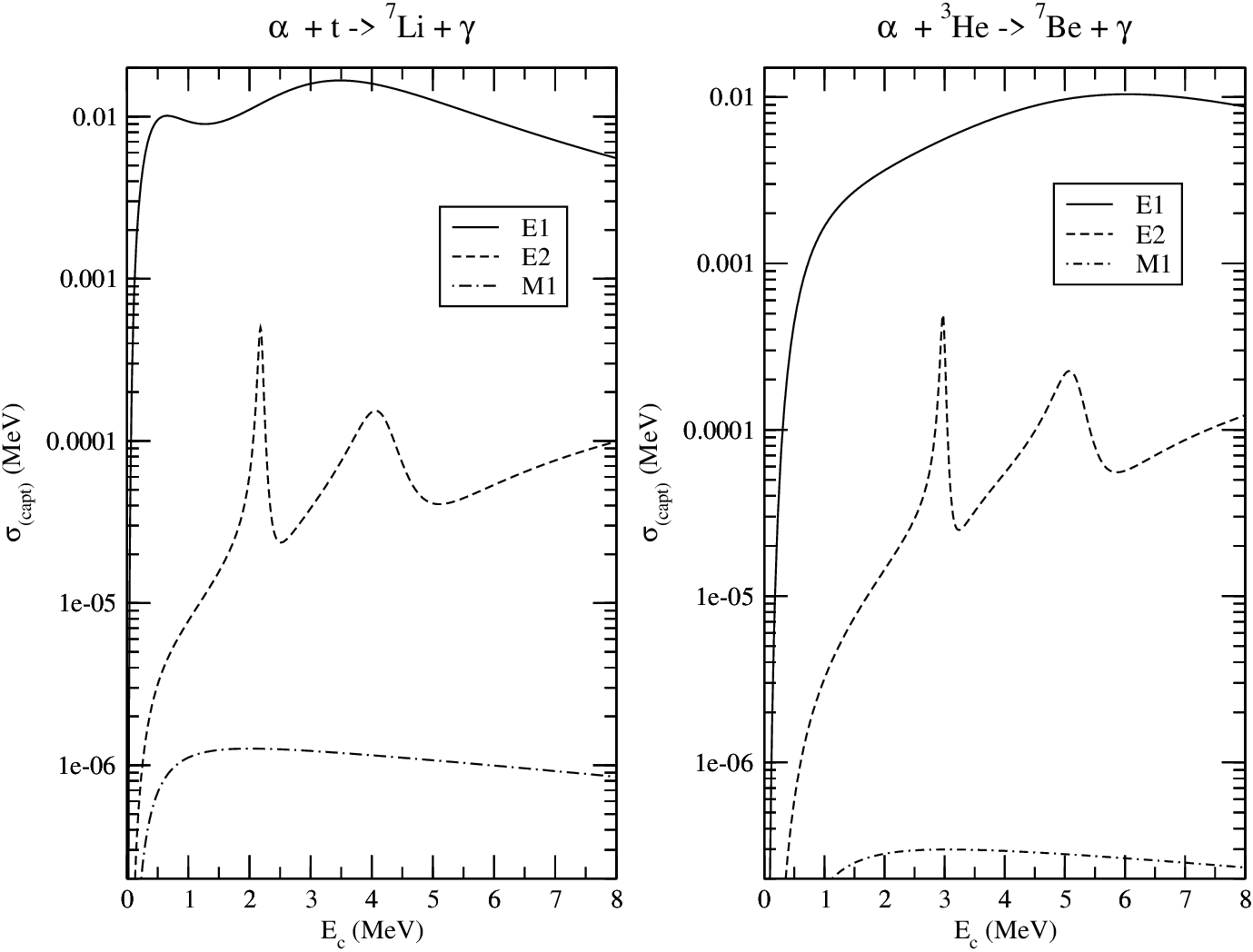}
\caption{Photo-dissociation and radiative capture processes (left pictures) and cross-sections 
(right) plotted against the continuum energy for reactions involving $^7$Li and $^7$Be. 
The E1,E2 and M1 contributions are shown separately and the total practically coincides with 
the electric dipole contribution. The E2 resonances are seen in each picture. The photo-dissociation 
is shown only from the ground state, but the radiative capture can go directly to the ground state or proceed 
through the first excited state.}
\label{cro}
\end{figure*}

The electric and magnetic response functions obtained so far can be used to calculate the cross sections for these two 
processes. The radiative capture cross-section for type ($\pi$ = Electric or
Magnetic) and multipolarity $\lambda$, can be expressed \cite{tb} as 
$$\sigma_{capt}(\pi\lambda,E_{C})= $$
\begin{equation}
=\frac{2(2j_{}+1)}{(2j_{\alpha}+1)(2j_{cl}+1)}
\left(\frac{k_{\gamma}}{k_{\alpha -cl}}\right)^2\sigma_{ph.dis}(\pi\lambda,E_{c})
\end{equation}
where $k_{\alpha -cl}$ is the wave-number for the relative motion of the two clusters (one is always an $\alpha$ 
particle in the present case) and $\frac{2(2j_{}+1)}{(2j_{\alpha}+1)(2j_{cl}+1)}$ is a spin factor 
from the detailed balance principle. This expression relies on the knowledge of the photo-dissociation cross-section 
that may be expressed as follows: 
\begin{equation}
\sigma_{ph.dis}(\pi\lambda,E_{C})=(2\pi)^3\frac{(\lambda+1)}{[\lambda(2\lambda+1)!!]^2}
k_{\gamma}^{2\lambda-1}\frac{dB(\pi\lambda,E_{c})}{dE_C}
\end{equation}
where $\frac{dB(\pi\lambda,E_{c})}{dE}$ is the reduced transition probability (defined as in Ref. \cite{tb}),  $k_{\gamma}$ is the photon wave-number, 
$\lambda$ is the multipolarity of the transition, $E_{c}=\hbar^2k_{\alpha -cl}^2/2\mu_r$ is the relative energy 
of the two clusters in the continuum and $E_{\gamma}=E_{c}+E_{b}$ is the photon energy. 
Here we have used the notation $\mu_r$ for the reduced mass and $E_b$ 
for the binding energy.

To evaluate the total 
photo-dissociation cross-section, starting from a given bound state, one has to sum over all possible angular momentum
transfer and possible allowed spins of the final states in the continuum. In the calculation of the total capture 
cross-section, in addition, one has to include all the bound states.

Fig. (\ref{cro}) shows a schematic picture for the two processes (left side), 
 and the respective cross sections (right side). 
They are very similar at a qualitative level, but the different binding energies of the ground states (mostly due 
to the different charge) makes up for changes in the position of the resonances and in the intensity of the transition
to the low-lyng continuum. 

This is evident from the comparison of fig. \ref{BeE1} of the present paper and fig. 1 of 
our older work on $^7$Li \cite{Forvit}. 
The BE(1) to low-lying continuum states has slightly different contribution from the $s$ and $d$ states
that explains the differences in the cross-section at small energies.
Indeed it is known that the shape and position of the maxima in the response are affected by binding energy, initial and final angular momenta and Coulomb barrier height \cite{tb,naga}.

Electric dipole clearly dominates the cross-section, although the electric 
quadrupole shows peaks due to the presence of two low-lying resonances. 
The electric quadrupole and the magnetic dipole are of comparable size and are nevertheless not negligibly small in 
comparison with the other contributions. 

The photo-dissociation cross-section is shown only from the ground state, but the radiative capture 
can go directly to the ground 
state or proceed through the first excited bound state. In the latter case, due to the small 
energy difference, the cross 
sections to the $p_{1/2}$ and $p_{3/2}$ are practically proportional and the proportionality is fixed 
by the phase space factor depending on the total spin, $j$.

\section{S-factors}

\begin{figure*}[t!]
\begin{center}
\includegraphics[clip=,width=.65\textwidth]{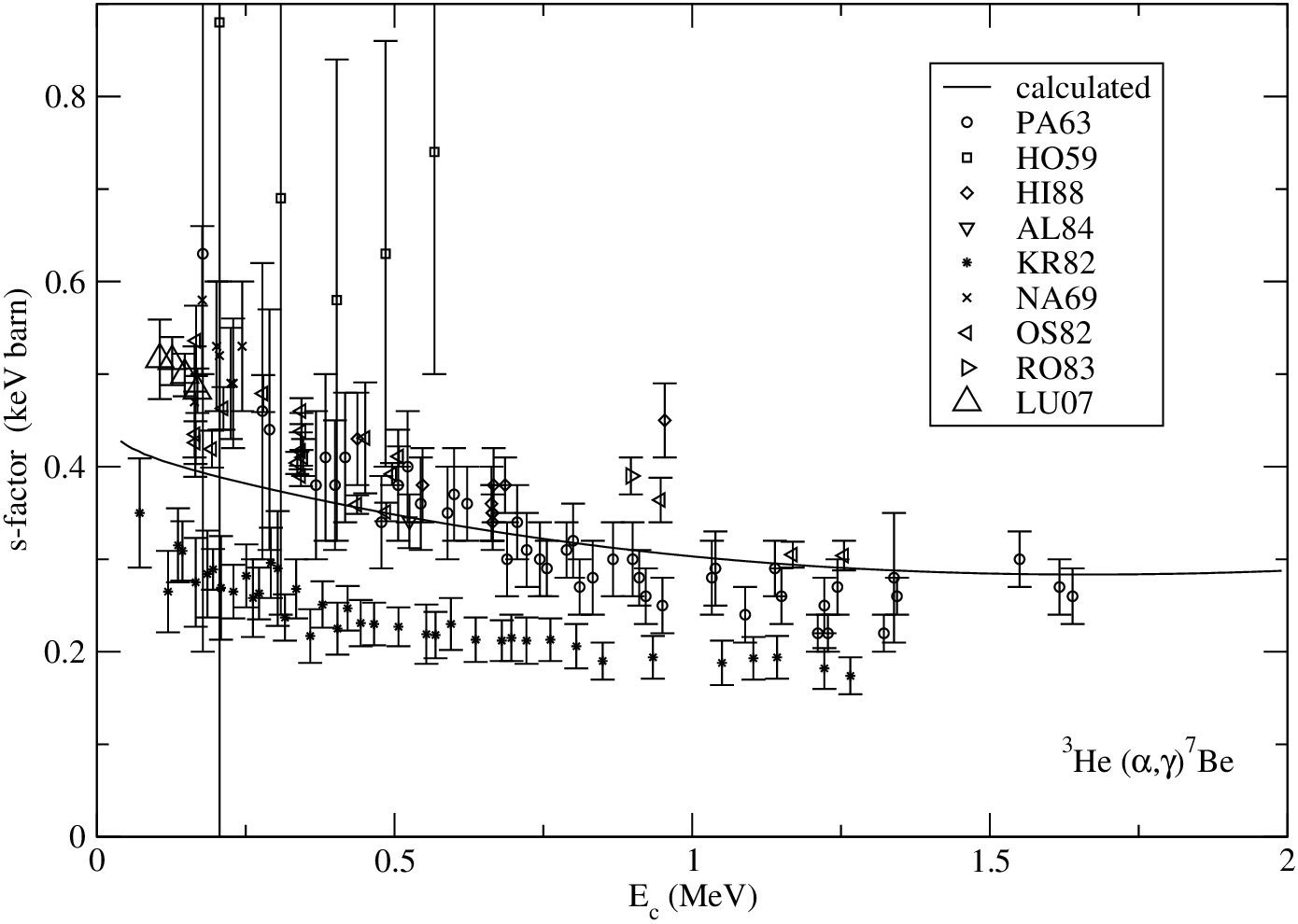}
\end{center}
\caption{S-factor for the $^3$He$(\alpha,\gamma)^7$Be reaction.}
\label{sfac-Be}
\end{figure*}

\begin{figure*}[t!]
\begin{center}
\includegraphics[clip=,width=.65\textwidth]{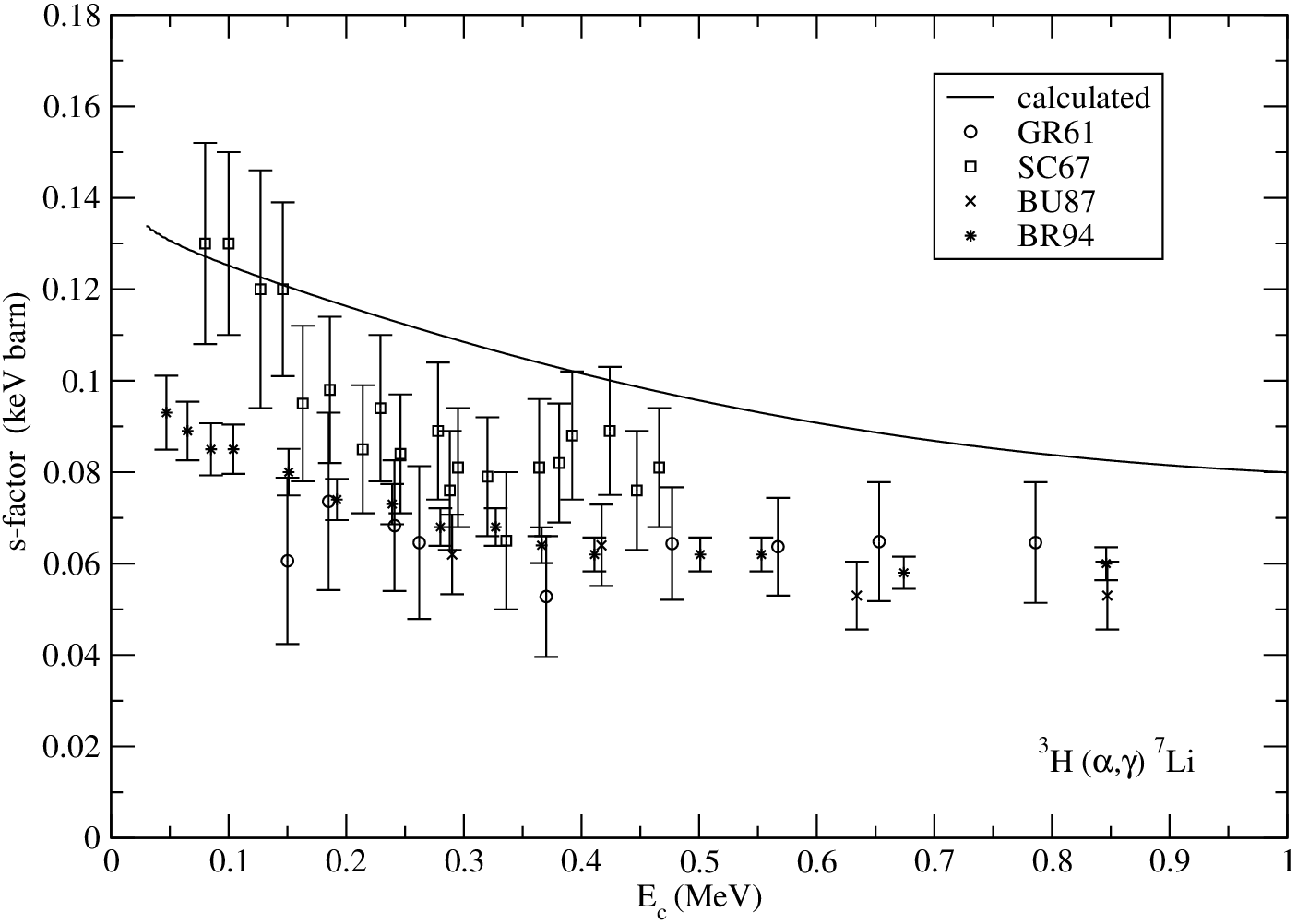}
\end{center}
\caption{S-factor for the $t(\alpha,\gamma)^7$Li reaction.}
\label{sfac-Li}
\end{figure*}

In the stellar environment the occurrence of high temperatures and the large density of 
light isotopes makes the $\alpha-$capture a very likely process. Tritium and $^3$He 
participate in $(\alpha,\gamma)$ reactions forming mass $A=7$ isobars, which are one 
possible getaway toward heavier nuclei. It is therefore important to understand well 
these processes and to have simple models to treat the reactions. Once again the 
dicluster picture qualitatively and quantitatively entails the correct
physics ingredients: indeed the astrophysical S-factor (a reparameterization of the cross-section, defined by 
$S=e_{rel}\sigma_{capt}(e_{rel})e^{2\pi\eta}$, 
where $\eta$ is the Sommerfeld parameter) for the reaction $^3$He($\alpha,\gamma$)$^7$Be is reasonably well reproduced 
as can be seen in figures (\ref{sfac-Be}). Good results are also obtained for the partner 
reaction, $^3$H($\alpha,\gamma$)$^7$Li, as shown in (\ref{sfac-Li}). 
Each curve has both a positive and a negative aspect: while the result for Be
has a good magnitude, but a non-optimal slope, the result for Li has a good
 shape, but it overestimates the data. Our results do not contain any rescaling of 
the S-factor or, in other words, we assume a spectroscopic factor for the 
$\alpha+$cluster configuration equal to one. This might be the reason 
for the observed deviation in the case of $^7$Li, where the 
$^6$Li$+$n component may affect the  total magnitude although possibly 
not the overall shape.
The reason for the flat shape of the $^7$Be curve might also reside in the 
prescription we have used to treat the dicluster system. Perhaps it is 
an indication that other components are affecting the properties at low 
energies although the results in table II were very satisfactory.

The various sets of 
experimental measurements that are reported in these figures together with the 
calculated values, are taken from the NACRE database \cite{NACRE}, where an extensive list of 
references can be found.
A simple parabolic fit of the model curves give a S(0) of 0.42 and 0.13 $keV\;b$ for beryllium and lithium respectively.

Other more refined approaches, like the 
R-matrix calculations (see discussion in Ref. \cite{Luna}), or the Multichannel Algebraic Scattering approach \cite{Cant} 
have been applied with success to the same reaction. Although our model is much simpler than the cited approaches, 
it has a comparable predictive power. Of course this might still not be enough for elaborated nucleosynthesis models that must
consider a network of nuclear reactions with very precise fits. In that case a more precise determination of the S-factor 
might be necessary and one must go beyond the simple cluster picture.

\begin{figure}[t!]
\begin{center}
\includegraphics[clip=,width=.48\textwidth]{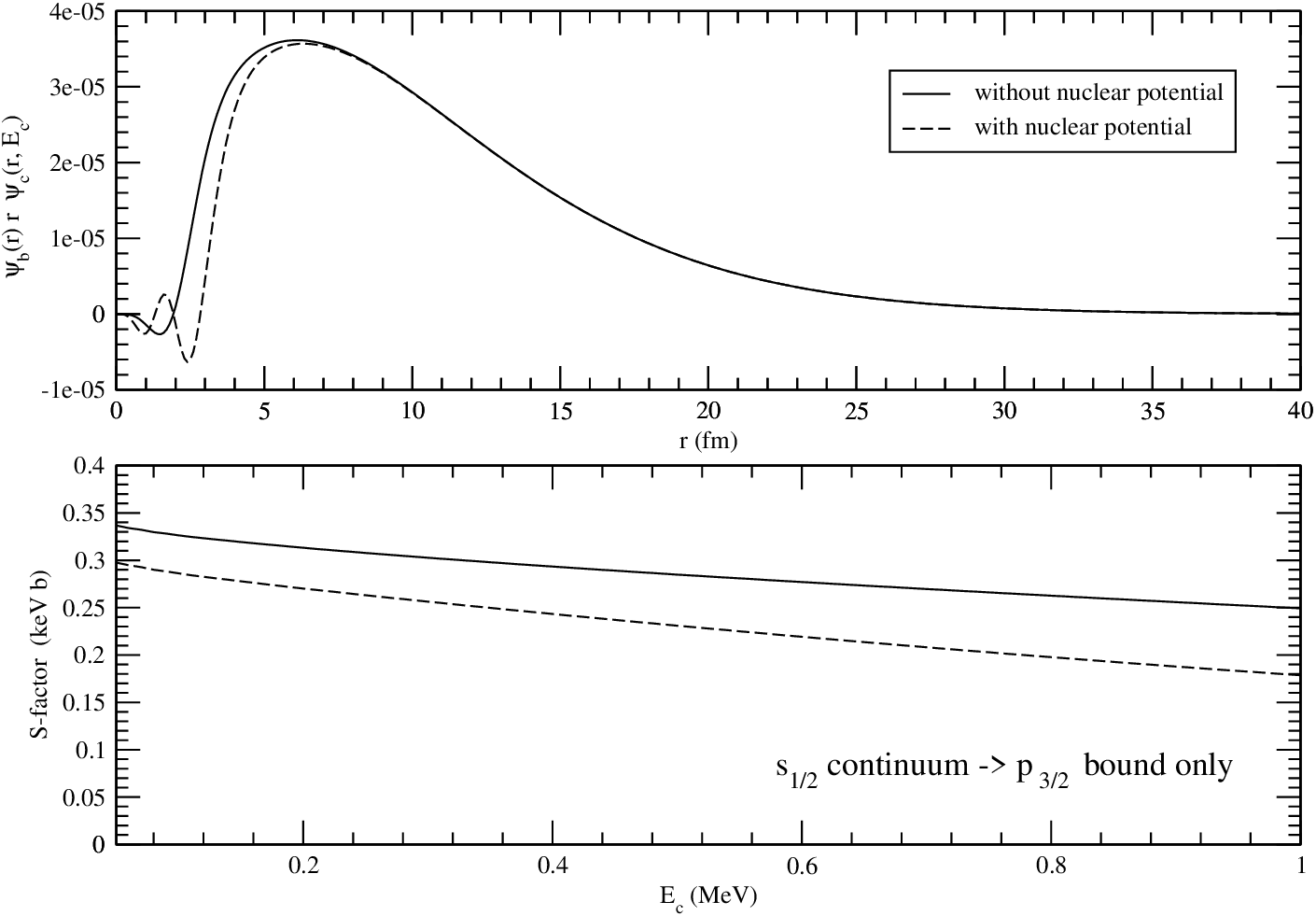}
\end{center}
\caption{Integrand of the response function for the transition s$_{1/2} \rightarrow$ p$_{3/2}$ in $^7$Be for a given energy $E_c=50$ 
keV (upper panel) and S-factor for the same transition in the low-energy regime. 
The two curves are calculated including (dashed) or excluding (solid) the nuclear potential 
in the evaluation of the continuum wavefunction.}
\label{nuc_imp}
\end{figure}

The whole idea of parametrizing the capture cross-section in terms of the S-factors is 
precisely to smooth out the effects of the Coulomb interaction 
(by means of the exponential term containing the Sommerfeld parameter). The important effect of the nuclear 
interaction in the low-energy regime can be seen in 
Fig. \ref{nuc_imp} and cannot be neglected.  
The upper panel displays the integrand of the response function as a function of $r$ for a given energy of 
$E_c=50$ keV in the continuum. We have selected the  s$_{1/2} \rightarrow$ p$_{3/2}$ transition in $^7$Be 
for the sake of simplicity.
Clearly the nuclear potential does not affect the tail of the wavefunction, that matches a pure Coulomb wavefunction at large
distances, but at distances smaller than the position of the Coulomb barrier, the presence of the nuclear potential 
affect the continuum wavefunctions and therefore the response in the continuum. The S-factor, in the low-energy regime, is 
displayed in the lower panel. The nuclear potential is seen to lower the S-factor significantly and cannot be excluded from 
calculations, as already found by several authors.

\section{Conclusions}

An elementary dicluster model is found to provide an effective description of static as well 
as dynamic properties of some light nuclei and can be used with confidence in reaction models. Although 
{\it ab initio} many-body models are expected to provide better results in the general case, we have shown 
that, for example, the astrophysical S-factor, which is a crucial quantity for interdisciplinary applications of 
nuclear physics, is reasonably well reproduced by the dicluster model.

In particular we have investigated the magnetic properties of $A=7$ isobars, introducing a new molecular magnetic 
dipole sum rule, that has a close analogy with the single particle magnetic sum rule \cite{Trai}, 
and finding that (as expected) the contribution to photo-dissociation and radiative capture processes is rather small. 
This is in contrast with the strong M1 photo-dissociation peak at low energies in the case of deuteron, that has 
motivated our investigation.

\section*{ Appendix: derivation of standard formulae for dicluster nuclei}

We summarize here a few important formulas valid for dicluster nuclei 
\cite{WalFli,Liu}. They are often used in the literature, but the derivation,
albeit elementary in some cases, is rarely reported.

\subsection*{Matter radius}
The mean square matter radius of a nucleus with mass number A is
defined as 
\begin{equation}   \label{dim1.1}
  r^2 = \frac{\sum_{i=1}^A {\vec{r}_i}^{~2}}{A}
\end{equation}
Assuming that the nucleus made up by two clusters with mass number A$_1$, A$_2$
and A=A$_1$ + A$_2$, Eq. (\ref{dim1.1}) becomes 
\begin{equation} \label{dim1.2}
  r^2_{A} = \frac{\sum_{i=1}^{A_1} (\vec{R}_{A_1} + \vec{R}_i)^2 
+ \sum_{j=1}^{A_2} (\vec{R}_{A_2}+ \vec{R}_j)^2}{A}
\end{equation}
where $\vec{R}_{i,j}$ are position vectors measured from the center of mass of each cluster,
A$_{1,2}$ respectively; $\vec{R}$ is the intercluster distance that can be split into two
vectors, $\vec{R}_{A_1}$ and $\vec{R}_{A_2}$ related by
\begin{eqnarray}   \label{separation}
\vec{R}_{A_1} &=& -\frac{A_2}{A}\vec{R} \nonumber \\
\vec{R}_{A_2} &=& \frac{A_1}{A}\vec{R}
\end{eqnarray}
By inserting (\ref{separation}) in (\ref{dim1.2}) and upon taking the expectation values rather than the operators, one gets
\begin{equation} \label{r2AB}
\langle r^2\rangle_{A_1+A_2} = \frac{A_1}{A} \langle r^2\rangle_{A_1} + \frac{A_2}{A}\langle r^2
\rangle_{A_2} + \frac{A_1A_2}{(A)^2}\langle R^2\rangle 
\end{equation}

\subsection*{Charge radius}
The mean square charge radius of a nucleus with atomic number Z is 
\begin{equation} \label{dim2.1}
 r^2_{ch} = \frac{\sum_{i=1}^Z \vec{r}_i^{~2}}{Z}
\end{equation}
For a nucleus composed by two clusters with atomic numbers 
Z$_1$, Z$_2$ and Z=Z$_1$+Z$_2$ eq. (\ref{dim2.1}) becomes 
\begin{equation} \label{dim2.2}
r^2_{ch} = \frac{\sum_{i=1}^{Z_1} (\vec{R}_{A_1} + 
\vec{R}_i)^2+ \sum_{i=1}^{A_2}(\vec{R}_{A_2}+\vec{R}_i)^2}{Z}
\end{equation}
Using (\ref{separation}) and taking the expectation values of operators, one finds
$$\langle r^2 \rangle_{A_1+A_2}^{ch} = \frac{Z_1}{Z} \langle r^2
\rangle_{A_1}^{ch} + \frac{Z_2}{Z} \langle r^2 \rangle_{A_2}^{ch} +  $$
\begin{equation} \label{r2ch}
\frac{
\langle R^2 \rangle}{Z} \Big( Z_1\cdot \Big(\frac{A_2}{A}\Big)^2 + Z_2\cdot\Big(
\frac{A_1}{A}\Big)^2\Big)
\end{equation}

\subsection*{Magnetic dipole moment}
The magnetic dipole moment operator is defined as
\begin{equation} \label{mucl}
\vec{\mu} = \sum_{i=1}^{A+B} \mu_i = \sum_{i=1}^{A+B} \left( g^l_i \vec{l}_i + 
g^s_i \vec{s}_i \right) \mu_N
\end{equation}
with $g^l=1$ for protons, $g^l=0$ for neutrons,  $g^s=5.58$ for protons and  
$g^s=-3.83$ for neutrons. The orbital angular momenta are defined as 
$\vec{l}_i=\vec{r}_i \times \vec{p}_i$, where
$\vec{r}_i=\vec{R_{A_1}} +\vec{R}_i$ for cluster A$_1$ and $\vec{r}_j=\vec{R_{A_2}} 
+\vec{R}_j$ for cluster $A_2$. The same rule can be applied to momenta 
$\vec{p}_i$. Labelling with $\vec{\mu}_1, \vec{\mu}_2$ 
 the intrinsic magnetic dipole moments of each of the two clusters,
(\ref{mucl}) becomes
\begin{equation}  \label{muAB}
\vec{\mu} = \vec{\mu}_1 + \vec{\mu}_2 +\mu_N \underbrace{\frac{Z_1A_2^2 + Z_2A_1^2}{AA_1A_2}}_{G}\vec{L}
\end{equation}
where one has isolated in the third term the contribution from the relative orbital angular momentum, $\vec{L}$.
The total magnetic moment is therefore simply the sum of the magnetic moments of the two subsystems plus the 
contribution from the cluster-cluster orbital motion, which vanishes if the two cluster are in a relative s-state.

\subsection*{Matter and charge quadrupole moments}
Matter and charge quadrupole operators 
moments are defined as
\begin{eqnarray} \label{dim5.1}
Q^{matter}&=&  \sum_{i=1}^{A_1+A_2} r_i^2 Y_{20}(\theta_i,\phi_i) \\
Q^{charge} &=& \sum_{i=1}^{A_1+A_2} e_i  r_i^2 Y_{20}(\theta_i,\phi_i)
\end{eqnarray}
and, by using the definition of the spherical harmonic and 
by splitting the sum over the total nucleon number into two 
sum over the nucleons of each 
cluster, we obtain two relationships that contain the 
intrinsic quadrupole moments of the clusters and a term 
that expresses the quadrupole 
operator for the relative motion, namely:
\begin{eqnarray} \label{dim5.2}
Q^{mat}&=& Q^{mat}_{A_1} + Q^{mat}_{A_2}+\frac{A_1A_2^2 + 
A_2A_1^2}{A^2} R^2 Y_{20}(\Theta,\Phi) 
\nonumber \\
  Q^{ch} &=& Q^{ch}_{A_1} + Q^{ch}_{A_2}+ \frac{Z_1A_2^2 + 
Z_2A_1^2}{A^2} R^2 Y_{20}(\Theta,\Phi) 
 \nonumber \\
\end{eqnarray}
With these operators one can calculate the quadrupole moments by evaluating 
the expectation value in the ground state with the maximally aligned 
magnetic substate:
\begin{equation}
Q^{mo} = \sqrt{\frac{16\pi}{5}}\langle L, M=L \mid Q \mid L, M=L \rangle
\end{equation}
For example in the case of the matter quadrupole moment one has:
$$Q^{mat.mo} = Q^{mat.mo}_{A_1} + Q^{mat.mo}_{A_2}+ $$
\begin{equation}
+\frac{A_1A_2^2 + 
A_2A_1^2}{A^2} \langle R^2 \rangle 2(2L+1)(-1)^L  
\left( {L \atop -L} {2 \atop 0} {L \atop L}\right) 
\left( {L \atop  0} {2 \atop 0} {L \atop 0}\right)
\end{equation}
where $L$ is the angular momentum of the relative motion.
In the case of the two mirror isobars lithium and beryllium,
we have $L=1$ and therefore the above formula reduces to 
\begin{equation}
Q^{mat.mo} = Q^{mat.mo}_{A_1} + Q^{mat.mo}_{A_2}-\frac{2}{5}\frac{A_1A_2^2 + 
A_2A_1^2}{A^2} \langle R^2 \rangle \;.
\end{equation}
Analogous expressions may be derived for the charge quadrupole moments that
differ only in the coefficient that depends linearly on charges rather than masses.
Notice also that, while the coefficient of the relative motion part 
for the charge quadrupole moment is different for lithium and beryllium, 
the one for the matter quadrupole moment is the same. Another difference 
comes from the average square radius of the relative motion wave function, 
which is not identical for the two nuclei.

\end{document}